# Title: On the emergence of agency

**Authors:** Aliza T. Sloan[1]*, J. A. Scott Kelso[1,2]

**Affiliations:**

[1]Center for Complex Systems & Brain Sciences, Florida Atlantic University, FL, USA.
[2]Intelligent Systems Research Centre, Ulster University, Derry~Londonderry, N. Ireland

*Corresponding author e-mail: asloan2014@fau.edu

**Abstract:** How do human beings make sense of their relation to the world and realize their ability to effect change? Applying modern concepts and methods of coordination dynamics we demonstrate that patterns of movement and coordination in 3-4 month-olds may be used to identify states and behavioral phenotypes of emergent agency. By means of a complete coordinative analysis of baby and mobile motion and their interaction, we show that the emergence of agency takes the form of a punctuated self-organizing process, with meaning found both in movement and stillness.

**One-Sentence Summary:** Revamping one of the earliest paradigms for the investigation of infant learning, we show that agency's emergence can be explained as a bifurcation or phase transition in a dynamical system that spans baby, brain and environment.

**Main text:**

How does *agency*, action towards an end, emerge in humans? How might it be understood? Answering these questions requires a clearly defined experimental model system. A possible entry point is the Mobile Conjugate Reinforcement (MCR) paradigm (*1*) which has been used to investigate the nature of infant motor development, learning and memory (*2-5*). Only recently have quantitative models of MCR appeared (*6-7*), which, along with new ideas about the nature of cognition as an embodied, embedded and extended process (*3, 8-9*) force a re-examination of the phenomenon itself. In MCR, when one of the infant's feet is tethered to a mobile, kick rate increases. Although it is assumed that mobile and baby interact, understanding the nature of their relationship requires a quantitative dynamical analysis that covers the baby's movements, the mobile movement and the coordination between them. We show here that the MCR phenomenon when fully analyzed, supports the theory that human agency is a self-organized dynamical process (*6-7, 10*).

Quantitative modeling of agency's emergence relies on elucidating the temporal evolution of baby~mobile coupling. At some point during infant~mobile interaction, the baby may realize the two aspects are causally coupled (*6, 10*), his/her perceived control over the mobile becoming a main driver of action (*11-15*). Anecdotal evidence suggests that causal powers are discovered in coordination (*16*), but up to now quantifiable signatures of such are wanting (*12, 17*). Conscious agency, for instance, may appear suddenly as an 'aha!' experience (*10, 18*) or gradually, but only a complete dynamical analysis of the experimental arrangement can tell.

Here we quantify for the first time, the organism (baby)~environment (mobile) relation and how it evolves in time. In doing so, we elucidate the nature of the dynamics from spontaneous to intentional action and highlight several mechanisms overlooked in MCR. Among these are infant freezing on first exposure to mobile movement (but see *19*), the informative nature of infant pauses, and continued kicking after the baby is decoupled from the mobile (*11*) --strongly suggesting the baby is anticipating the sensory consequences of self-produced mobile movement.

Whether infant behavior during coupled interaction reflects a realization of causal connection or a generalized response to raw stimulation from a moving mobile can be clarified by including a condition wherein the mobile moves independently from the infant. Therefore, we tracked foot and mobile motion in 3D space at 100 Hz in 16 3-4-month-old infants across four experimental stages: **1)** A **spontaneous baseline** in which the mobile does not move. In the absence of stimulus motion, infant activity is spontaneously generated; **2)** Next, a **non-contingent baseline** measures infant reaction to experimenter-triggered mobile motion; **3)** A **tethered phase** which translates movements of the *trigger foot* (connected to a sensor by two ribbons) into mobile rotation rate; **4)** Finally, the ribbons are detached, and mobile is stationary during an **untethered phase**. (Fig. 1)

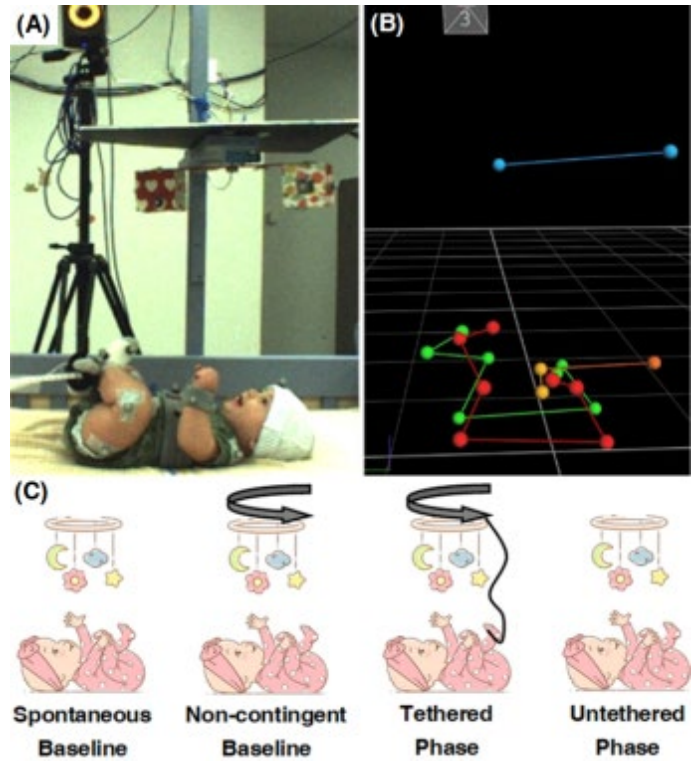

**Fig. 1. Experimental setup. (A)** Infant and mobile are connected via ribbons and outfitted with spherical motion markers, **(B)** with marker positions represented in 3D space (infant's left: red, right: green, torso/head: orange, mobile: blue). **(C)** The experimental paradigm proceeds from left to right (see *S1*).

Under which conditions do infants transition from being an observer to an active agent? We characterized multidimensional coordination between the feet and between each foot and the mobile for all infants across the experiment. As expected, the feet were more coordinated with each other ($r = .56$) than with the mobile during non-contingent baseline (trigger foot~mobile coordination: $r = .24$, $p < .001$; unconnected foot~mobile coordination: $r = .20$, $p < .001$). However, during tethering, trigger foot~mobile coordination ($r = .73$) was far stronger than either unconnected foot~mobile ($r = .52$, $p < .001$) or inter-foot coordination ($r = .46$, $p = .001$). (See *Table S1*.)

How does infant behavior change across conditions? Unexpectedly, infants moved *less* when the mobile began moving. Foot movement dropped 34% during the non-contingent baseline and remained suppressed in the first minute of tethering ($p < .001$), challenging the widely held assumption that mobile motion rewards/stimulates kicking. The mobile's initially suppressive effect also calls into question standard cut-offs of contingency detection, classically defined as 150% increase in foot activity during tethering compared to spontaneous baseline (*1*). Infants who substantially increase activity during tethering but do not meet the standard criterion because they begin in a dampened state are dismissed, likely contributing to the high failure rates common in MCR (*2*). Indeed, infant movement nearly doubled on average during the most active minute of tethering relative to non-contingent baseline and the first minute of tethering. Notably, the peak tethered rate did not change significantly even after infants were disconnected ($p = .19$), further challenging the idea that mobile activity reinforces infant activity (*17*). Instead, elevated movement rates after disconnection appear to reflect infant expectation of mobile response and attempts to reestablish the lost relationship. Peak activity rate during tethering was more strongly correlated to infants' activity in *reaction* to non-contingent mobile movement ($r = .64$, $p = .02$) than to their spontaneous rates ($r = .46$, $p = .07$), again underlining the importance of functional context for interpreting infant behavior.

We searched for signatures of agentive discovery in infants who increased activity by at least 150% from the first minute of tethering to their most active minute. Some infants might steadily increase movement rate during tethering, reflecting basic sensitivity to contingency, whereas others might realize their control over the mobile, suddenly increasing activity upon discovery (*10*). Trigger foot 3D displacement was differentiated in 1-minute windows to calculate tethered movement rate (*velocity*) and differentiated again to locate peak changes in movement rate (*acceleration*) (*e.g.*, Fig 2A).

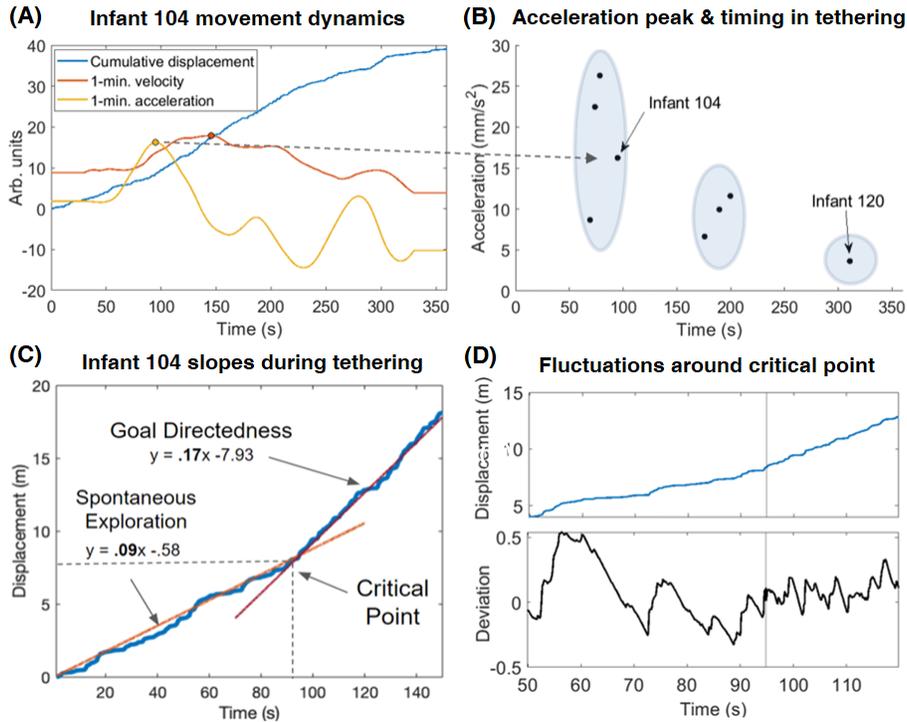

**Fig. 2. Individual phenotypes of emerging agency during tethered phase. (A)** Tethered foot displacement (blue), *velocity* (red) and *acceleration* (yellow) for infant 104 (measures scaled to fit; peak values marked as dots). **(B)** Magnitude versus timing of peak acceleration plotted for infants who increased activity ≥150% during tethering. Three distinct phenotypic patterns suggest different paths and timelines for agentive discovery. **(C)** Trigger foot cumulative displacement (blue) for infant 104 linearly modeled in the minute preceding and following peak acceleration identified in (A) reveals a critical transition point. Slope of displacement nearly doubles as infant shifts from exploring to purposefully triggering mobile movement. **(D)** Focusing on activity immediately surrounding transition (top panel; vertical line denoting time of transition), fluctuations (bottom panel) are greater before transition than after, a hallmark of nonequilibrium phase transitions (*20-23*).

Three clusters of infants emerged: infants whose rate increase peaked early, midway or late in tethering (Fig. 2B) with average rate increases of 281%, 175% and 151%, respectively. Timing of peak rate increase was inversely related to magnitude of change and total increase across tethering. Using traditional methods, infants 104 and 120 would be identically classified as having met the standard criterion for contingency detection. However, unlike infant 104 who quickly discovered his causal abilities, doubling activity over just one minute of tethering, infant 120 was still exploring her functional relationship with the mobile six minutes into tethering, slowly increasing activity throughout. Dynamics proves to be a critical tool for identifying moments of agentive discovery, differentiating agentive states and exposing underlying mechanisms. The finding of three distinct clusters of infants suggests that behavioral phenotypes of agentive discovery exist—and that dynamics provide a means to identify them.

To explore one form of phenotypic expression, infant 104's journey to agentive discovery was analyzed in detail (Fig. 2A, C-D). When quantified as a change in slope of displacement, the activity burst indicating agency's emergence resembles the sudden transitions seen in adult studies of sensorimotor learning (*23*) (Fig. 2C). Differentiating agentive states is a prerequisite to understanding rules for transitioning between them. The magnitude of deviation from the regression line is larger before the moment of discovery than after (Fig. 2D). Whereas phase transitions are typically anticipated by growing instability (*critical fluctuations*) and longer recovery times (*critical slowing down*), the magnitude of fluctuations in this infant progressively shrank. This unexpected *reduction* of fluctuations and *critical speeding up* just prior to transition reflects a decrease in exploratory behavior and an apparent increasing certainty of becoming a mindful body (*24*). Critically, infant 104's realization of self-agency occurred within the context of tight trigger foot~mobile coordination and after disruption to the unconnected foot~mobile relationship (Figs. 3B). When a strong, new relationship formed between one of the infant's limbs and the mobile, a less functionally relevant relation within the infant's body destabilized
(Fig. 3B).

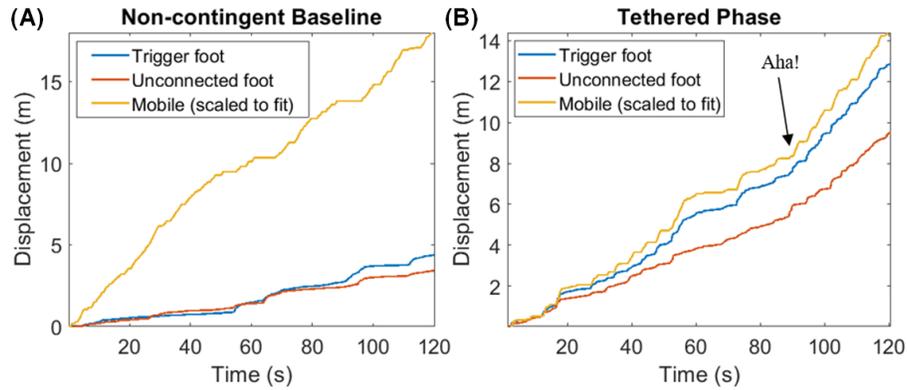

**Fig. 3. Changing coordination within the baby's body and between baby and mobile precede discovery. (A-B)** Displacement of infant 104's feet and the mobile demonstrate that the feet are more strongly related to each other than the mobile during baseline, whereas during tethering the trigger foot and mobile are tightly coupled and the relationship between the feet loosens after ~50s.

The influence of mobile activity on infant movement reveals further insight into the mechanisms of discovery. Infant 104 moved at some intrinsically preferred rate during spontaneous baseline (Fig. 4A, left). At the start of tethering, he began to move as before, but because movement now triggered mobile motion, 104 quickly froze and did not recommence until the mobile stopped (Fig. 4A, right), producing a bursting, staccato pattern of sharp starts and stops unique to the tethered phase. Thus, agentive realization (indicated by the activity spike at ~90s) appears to be directed by two opposing forces during tethering: an intrinsic drive to explore accompanied by a freezing response to novel mobile movement. Whereas episodes of movement communicate the degree of relatedness between infant and mobile, periods of inactivity afford crucial counterevidence. If the infant moved ceaselessly, there would be no way to rule out the possibility that some external force drives the mobile. Information of agency exists both in stillness and movement. For several babies, a sudden increase in trigger foot activity during tethering (*i.e.,* agentive realization) was preceded by a brief period of relative rest (Fig. 4B). After a short pause at ~170s in tethering, infant 104 further discovered how to reliably elicit ~2 mobile rotations per bout of movement (Fig. 4A). Such discoveries are made in the gaps.

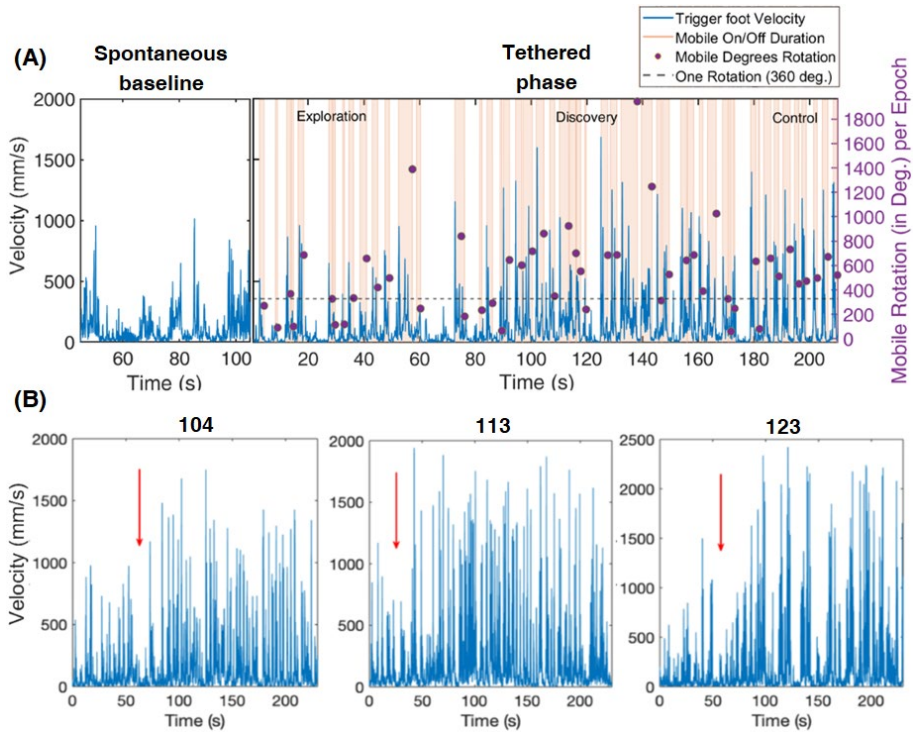

**Fig. 4. Agentive discovery is a punctuated process. (A)** Infant 104's 3D trigger foot velocity patterns (blue) changed drastically as a function of mobile activity (duration: width of orange bars, magnitude: dots). During spontaneous baseline (left), the infant made long bouts of slow, meandering movements. Momentary freezing occurred as infant attended to contingent mobile motion producing a bursting velocity pattern unique to the tethered phase (right). Alternating behavior generated many opportunities to explore effects of *both* self-movement and inactivity. After a lengthy pause, activity spiked at ~90s and mobile pauses progressively shrank as infant exercised newfound causal abilities. Now, each bout of activity elicited ≥1 mobile rotation. **(B)** For several infants, a period of relative rest (arrow) preceded a burst of trigger foot activity suggestive of sudden discovery.

Previous MCR studies collapsed environmental stimulus and organismic response. Now giving each aspect its due, we find that the emergence of agency can be explained as a bifurcation or phase transition in a dynamical system (*6*): from a less correlated state to a state where both movements of mobile and the tethered limb are highly coordinated. Departing from classical reinforcement frameworks, the present analysis of coordination dynamics shows that the emergence of agency is a punctuated, self-organizing process (*10*) with meaning found both in movement and stillness (*24*). That the mobile moves when the baby moves but not when the baby pauses—like Cezanne's pauses between brush strokes—confirms to the baby that 'I can make things happen.'

**Acknowledgments:**

**Funding:**

Work supported by the FAU Foundation

**Author contributions:**

Conceptualization: JASK, ATS

Methodology: ATS, JASK

Investigation: ATS, JASK

Writing—original draft: ATS, JASK

Writing—review and editing: ATS, JASK

**Competing interests:**

None.

**Data and materials availability:**

Data, including video recordings of study sessions, are stored securely in accordance with IRB rules and regulations. Data which can be de-identified can be shared upon request.

# Supplementary Materials for

## On the emergence of agency

Aliza T. Sloan, J. A. Scott Kelso

Correspondence to: asloan2014@fau.edu

**This PDF file includes:**

    Materials and Methods
    References (*25-27*)
    Table S1

**Materials and Methods**

<u>Recruitment</u>

Birth records were obtained from the Florida Department of Health. Infants were recruited by postcards which were sent to all households within a 1-hour drive of the university lab who had full-term (>36 weeks), healthy (Apgar >7) infants under 3 months of age. Infant gestational age was obtained from parents before testing to confirm that infants were full term. Informed consent was also obtained from parents before testing. This study was approved by the Florida Atlantic University Social, Behavioral and Educational Research Internal Review Board and the Florida Department of Health Internal Review Board.

<u>Demographics</u>

Sixteen 3-4-month-old full term infants (11 male, 5 female) participated in this study. Infants were on average 100.33 days old ($SD$ = 15.57 days). Using WHO growth charts, these infants were on average at the 60th percentile for weight given their age ($SD$ =22.52%). INFANIB assessments found no significant motor development delays.

<u>Apparatus</u>

The current study measured infant and mobile movement in 3-dimensional space using Vicon motion capture technology which employs 8 infrared cameras. The system was set up to capture marker positions (distance from origin of the 3D space) at a rate of 100 Hz. Infants interacted with a feedback system which translated infant leg movements into mobile rotation. The apparatus consisted of a wooden arm with a pivot joint at its midpoint suspended above the crib parallel to the length of the crib. This arm was easily rotated in one plane only. An Adafruit BNO055 absolute orientation sensor, equipped with a 9 degree-of-freedom accelerometer, magnetometer and gyroscope was attached to one end of the rotating arm suspended above the crib. This side of the arm was also very slightly weighted so that at rest the end with the sensor pointed downward. Two ribbons were tied to the other end of the arm. When these ribbons are pulled downward, the arm rotates and the sensor rises upward. The other ends of the ribbons were snapped onto a sock placed on the infant's foot. Euler angles of the arm were measured by the sensor and transmitted to an Arduino Uno board which calculated the change in angle from cycle to cycle. If the change in Euler angle was greater than 1 degree, the Arduino sent a signal to a DC motor to begin spinning. A simple mobile consisting of a wooden plank with colorful cubes on either end was attached to the rotating shaft of the DC motor and was suspended above the infant's face. The magnitude of the change in angle of the sensor was mapped onto the speed of the motor, meaning that a greater change in angle (associated with a leg movement of greater amplitude and force) produced faster mobile rotation. When the change in Euler angle was less than one degree, the motor was not activated. After measuring the angular change of the bar at the end of movements, the one-deg. threshold was chosen to be just high enough to eliminate the effects of the bar's post-kick wobble on the movement of the mobile. A piece of foam board was placed above the mobile to both focus the infant's attention on the mobile and block the movement of the rotating arm outfitted with the accelerometer from the infant's view.

Procedure

The experiment took place in the Human Brain and Behavior Laboratory at Florida Atlantic University. During the lab visit, the infant was outfitted with socks and reflective markers and then placed supine into a crib.  The MCR procedure involved four phases: spontaneous baseline (no mobile movement), non-contingent baseline (experimenter triggered mobile movement), tethered phase (tethered foot triggered mobile response), and untethered phase (tethered foot is disconnected and mobile is stationary). Each phase lasted two minutes, with the exception of the

tethered phase, which lasted up to 6 minutes. If an infant began to cry and could not be quickly

soothed, the current phase of the experiment was ended prematurely. On average, the time

lengths for each phase were: spontaneous baseline - 101.53s ($SD$ = 42.94), non-contingent

baseline - 97.46s ($SD$ = 44.72 sec), tethered phase - 274.00s ($SD$ = 119.15) and untethered phase

- 95.80s ($SD$ = 43.21 sec). During the non-contingent baseline, care was taken to not coordinate

mobile motion with infant motion.

Data preparation

*3D Velocity & Displacement.*

The first derivative of the position (velocity) for each of the three axes was calculated using the

central difference method:

$$v_{a,n} = \frac{x_{n+1} - x_{n-1}}{2T},$$ (1)

where $a$ is equal to the axis, and $T$ is equal to the sampling period (1/100 s). $v$ is a 3 x $n$ matrix

(*25*). The magnitude of the 3D velocity is equal to the square root of the sum of the squared velocity matrix. The magnitude of the 3D velocity was calculated in this way for each foot.

Three-dimensional displacement was calculated by cumulatively summing the 3D velocity.

*Mobile Rotation & Epochs of Mobile Movement.*

For any episode of infant movement, the starting position of the mobile is arbitrary. Furthermore,

while the built-in Matlab function atan2.m can be used to convert position to angle, each time the mobile passes the point on its orbit which corresponds to 0 radians, atan2 inserts an artificial jump of $2\pi$ radians to signify the beginning of the next cycle. Mobile X and Y positions were transformed using trigonometric properties to calculate rotational angle of the mobile, ranging from 0 to $\pi$ radians. Mobile angle reset to 0 each time the mobile stopped moving. However, the pseudo-activity from the reset did not contribute to the present data analysis and was removed for more accurate visualization. Dividing the cumulative sum of the absolute change in angle by 2 across an interval provides the number of rotations in that time period. (Dividing by 2 is necessary as $2\pi$ radians = 1 rotation).

A filter was constructed to investigate whether the length of time that the infants kept the mobile in motion changed across the tethered phase. The filter identified bouts of mobile motion when mobile velocity exceeded 30 mm/s for at least 150 ms. Mobile epochs were defined as bouts of mobile motion that resulted in at least one-tenth of a full rotation. Epoch start and stop times were then retained for further analysis

*3D Movement Dynamics within the Tethered Phase.*

While it was predicted that infant activity would increase across the tethered phase, it was also logical to suppose that each infant would peak in activity at different time points during the tethered phase due to individual differences. Using a MATLAB function, movingslope.m, (*26*) a 1-minute-wide moving window was applied to the cumulative displacement curve of the trigger foot to estimate its rate of displacement (mm/s) across the Tethered phase using a locally linear approximation. If multiple maxima were found to exist, the timing of the earliest maximum was retained as it represents the point when the infant first reached its maximum rate. These data were used to identify the start time of the most active minute and the maximum rate of displacement for each infant during the tethered phase. The tethered phase peak rate does not represent a set point in time during the tethered phase across infants. Rather, it is a measure of activity across a time window comparable to other time samples (one minute) reflecting maximum activity during the tethered phase across infants. This step allowed peak rate to be compared to movement rates of other experimental phases. The MATLAB function was reapplied to the displacement rate output (1-min. velocity) to calculate acceleration ($mm/s^2$) in 1-minute windows. Peak acceleration is equivalent to the greatest change in rate of activity across one minute of the tethered phase. Acceleration values represent whether changes in activity during the tethered phase occur abruptly or gradually.

*Coordination Dynamics.*

Coordination between each infants' feet and between each foot and the mobile were calculated. Standard methods of coordination analysis in signal processing are cross-correlation and cross-covariance. Fisher transforms of normalized cross-covariance values produce Pearson product moment coefficients which can then be statistically compared (*27*). Here, the magnitudes of various multidimensional velocity vectors (*i.e.,* between the two feet, between the trigger foot and the mobile, between the unconnected foot and mobile) were cross-covaried using the aforementioned method to obtain Pearson r-values and tested using repeated measures ANOVAs.

**Table S1.**

Coordination of 3D Velocity (*r*-values)

| Coordinative Pair | Spontaneous baseline | | Non-contingent baseline | | Tethered phase | | Untethered phase | |
|---|---|---|---|---|---|---|---|---|
| | *M* | *(SD)* | *M* | *(SD)* | *M* | *(SD)* | *M* | *(SD)* |
| Trigger foot~Mobile | *na* | *na* | .24 | (04) | .73 | (.11) | *na* | *na* |
| Unconnected foot~Mobile | *na* | *na* | .20 | (.04) | .46 | (.12) | *na* | *na* |
| Inter-foot | .49 | (.13) | .56 | (.13) | .45 | (.17) | .45 | (.15) |